# Charge order landscape and competition with superconductivity in kagome metals

Mingu Kang[1,2*†], Shiang Fang[2*], Jonggyu Yoo[1,3*], Brenden R. Ortiz[4], Yuzki M. Oey[4], Jonghyeok Choi[1,3], Sae Hee Ryu[5], Jimin Kim[6], Chris Jozwiak[5], Aaron Bostwick[5], Eli Rotenberg[5], Efthimios Kaxiras[7,8], Joseph G. Checkelsky[2], Stephen D. Wilson[4], Jae-Hoon Park[1,3†] & Riccardo Comin[2†]

[1]Max Planck POSTECH/Korea Research Initiative, Center for Complex Phase of Materials, Pohang 790-784, Republic of Korea.
[2]Department of Physics, Massachusetts Institute of Technology, Cambridge, Massachusetts 02139, USA.
[3]Department of Physics, Pohang University of Science and Technology, Pohang 790-784, Republic of Korea.
[4]Materials Department, University of California Santa Barbara, Santa Barbara, California 93106, USA.
[5]Advanced Light Source, E. O. Lawrence Berkeley National Laboratory, Berkeley, California 94720, USA.
[6]Center for Artificial Low Dimensional Electronic Systems, Institute for Basic Science (IBS), Pohang 790-784, Republic of Korea.
[7]Department of Physics, Harvard University, Cambridge, Massachusetts 02138, USA.
[8]John A. Paulson School of Engineering and Applied Sciences, Harvard University, Cambridge, Massachusetts 02138, USA.
*These authors contributed equally to this work.
†Correspondence should be addressed to iordia@mit.edu, jhp@postech.ac.kr, rcomin@mit.edu.

**In kagome metals $A$V$_3$Sb$_5$ (A = K, Rb, Cs), three-dimensional charge order (3D-CO) is the primary instability that sets the stage for other collective orders to emerge, including unidirectional stripe order, orbital flux order, electronic nematicity, and superconductivity. Here, we use high-resolution angle-resolved photoemission spectroscopy to determine the microscopic structure of three-dimensional charge order (3D-CO) in $A$V$_3$Sb$_5$ and its interplay with superconductivity. Our approach is based on identifying an unusual splitting of kagome bands induced by 3D-CO, which provides a sensitive way to refine the spatial charge patterns in neighboring kagome planes. We found a marked dependence of the 3D-CO structure on composition and doping. The observed difference between CsV$_3$Sb$_5$ and the other compounds potentially underpins the double-dome superconductivity in CsV$_3$(Sb,Sn)$_5$ and the suppression of Tc in KV$_3$Sb$_5$ and RbV$_3$Sb$_5$. Our results provide fresh insights into the rich phase diagram of $A$V$_3$Sb$_5$.**

The family of $A$V$_3$Sb$_5$ is a newly discovered series of kagome-based compounds realizing unconventional many-body phases and nontrivial electronic topology (Fig. 1c,d)[1,2]. In close analogy with other strongly correlated systems such as Cu- and Fe-based high-temperature superconductors[3,4], as well as moiré superlattices[5], a cascade of coupled symmetry-broken electronic orders has been observed in $A$V$_3$Sb$_5$. These include translational symmetry breaking in the form of 2×2 charge order (CO) below $T_{CO} \approx$ 78~102 K[2,6], rotational symmetry breaking in the form of unidirectional 1×4 stripe order below $T_{SO} \approx$ 50~60 K[7-10], a time-reversal symmetry breaking orbital flux phase below $T_f \approx$ 70 K[11], and superconductivity below $T_c \approx$ 0.92~2.5 K[2,12,13]. Understanding the origin, nature, and interrelation between these electronic orders is a major frontier of this emerging research field.

Notably, previous theoretical and experimental investigations point toward the unique role of the electronic structure of the underlying kagome lattice in driving the rich physics of $A$V$_3$Sb$_5$[14-20]. In the ideal limit, the kagome lattice exhibits multiple singularities in its electronic dispersion, including Dirac fermions at the Brillouin zone corner $K$, van Hove singularity (vHS) at the zone edge $M$, and flat bands across the whole Brillouin zone (Fig. 1a). Depending on the band filling fraction $n$, these electronic states may engender various topological and correlated phases as extensively investigated for more than a decade[21-26]. Especially at the vHS filling fractions $n =$ 3/12 and 5/12 (Fig. 1a), the Fermi surface of the kagome lattice is perfectly nested by three symmetry-equivalent reciprocal lattice vectors $Q_1 = (0.5, 0)$, $Q_2$, and $Q_3$ (Fig. 1b). Combined with the high density of states at vHS, the nesting creates a diverging electronic susceptibility and sets the stage for pairing in multiple channels, and the subsequent emergence of charge/spin order and superconductivity[24-26]. The electronic structure of the $A$V$_3$Sb$_5$ series follows this script, with multiple kagome-derived vHS sharply aligned to the Fermi level ($E_F$)[16-20]. Accordingly, as displayed in Fig. 1e, density functional theory (DFT) calculation of phonon frequency reveals six unstable modes exactly at the $Q_1$, $Q_2$, and $Q_3$ in-plane wave vectors – three at $M$ ($k_z = 0$) and the other three at $L$ ($k_z = \pi$) – indicating that the pristine kagome structure is unstable toward the 2×2 charge distortion[27-29]. Combined with the experimental identification of the 2×2 CO[2,6], this suggests that the toy-model vHS physics of the ideal kagome lattice is indeed realized within the $A$V$_3$Sb$_5$ system.

Importantly, the contribution of $L$ phonons with nonzero out-of-plane momentum ($k_z = \pi$) indicates that the full description of CO in the $A$V$_3$Sb$_5$ series needs to go beyond the limit of two-dimensional kagome lattice. The three-dimensional nature of CO in $A$V$_3$Sb$_5$ was reported in early studies[17,30], supporting either twofold (2×2×2) or fourfold (2×2×4) $c$-axis modulations[31,32]. As we illustrate in Fig. 1f,g, $M$-point ($L$-point) phonons are associated to V-V bond distortions in-phase (out-of-phase) across neighboring kagome planes[28]. Then, depending on the possible $3Q$ combinations of $M$ and $L$ phonons, various microscopic 3D-CO structures can be realized in the $A$V$_3$Sb$_5$ series (Fig. 1h-m): Star-of-David or SoD (–$M$,–$M$,–$M$); Tri-Hexagonal or TrH ($M,M,M$); Alternating SoD and TrH ($L,L,L$); Staggered SoD (–$M$,–$L$,–$L$); Staggered TrH ($M,L,L$); and Staggered Alternating SoD and TrH ($M,M,L$). Going beyond these simple $3Q$ superposition, the free energy analysis indicates that the trilinear coupling of $M$ and $L$ phonon modes plays a central role in defining the exact 3D-CO structures and the leading instabilities (see Supplementary Discussion 1 for details).[28] For example, due to the coupling term $\gamma_{ML} \sum M_i L_j L_k$, the $LLL$ phase is invariably accompanied by $MMM$ or inverse $MMM$ distortions and becomes ($LLL + MMM$) phase. The $MML$ phase similarly acquires additional $M$ and $L$ phonon distortions from the $\gamma_M M_1 M_2 M_3$ and $\gamma_{ML} \sum M_i L_j L_k$ terms and, most importantly, does not appear as a leading instability of the free energy.[28] In the following, we will consider the five structures – $MMM$, inverse $MMM$, $MLL$, inverse $MLL$, and ($LLL+MMM$) – as candidate 3D-CO structures, which are identified as the leading instabilities of the $A$V$_3$Sb$_5$ system in Ref.[28].

We emphasize here that identifying the exact structure and symmetry of 3D-CO is of paramount importance for understanding the electronic symmetry breaking transitions and phase diagram of $A$V$_3$Sb$_5$. This is because CO formation has the highest energy scale in $A$V$_3$Sb$_5$ series ($T_{CO} \approx 78 \sim 102$ K) and thus defines the background symmetry under which other electronic phases emerge (stripe order, nematicity, flux phase, and superconductivity). For example, if CO crystallizes in the inverse $MLL$, $MLL$, or $MML$ phases (Fig. 1k-m), $C_6$-rotational symmetry is spontaneously broken, which would indicate a possible lattice origin of the unidirectional 1×4 stripe order and electronic nematicity emerging below $T_{CO}$[7–10]. Furthermore, the detailed structure of 3D-CO can critically influence the competition between charge order and superconductivity, as different parts of the Fermi surface will be affected by the various 3D-CO patterns in a qualitatively different way. Despite its importance, the detailed microscopic structure of 3D-CO and its composition dependence in the $A$V$_3$Sb$_5$ series have not been determined conclusively, with different approaches – X-ray diffraction[31,32], coherent phonon spectroscopy[33], nuclear magnetic/quadrupole resonance[34], scanning tunneling microscopy simulations[27], and DFT[27,35] – producing divergent results.

In the present study, we establish the microscopic structure of 3D-CO and its evolution in the $A$V$_3$Sb$_5$ series by analyzing the detailed reconstruction of the electronic bands induced by 3D-CO. Using high-resolution angle-resolved photoemission spectroscopy (ARPES), we observe an unusual energy splitting of the kagome-derived vHS and Dirac bands, which is a direct consequence of the unit cell reconstruction in the 3D-CO phase. Crucially, the precise nature of the band splitting is highly sensitive to the intra-unit-cell stacking between different 3D-CO modulation patterns across adjacent kagome planes, which allows us to constrain the 3D-CO structure and symmetry in the $A$V$_3$Sb$_5$ series. Using this approach, we determine that the band splitting of CsV$_3$Sb$_5$ is most consistent with the Alternating SoD and TrH structure (*LLL+MMM* phase), while those in KV$_3$Sb$_5$, RbV$_3$Sb$_5$, and Sn-doped CsV$_3$Sb$_5$ are markedly different from the CsV$_3$Sb$_5$ and can be assigned to the staggered TrH structure (*MLL* phase). The tunability of the 3D-CO structure discovered here has important implications for superconductivity in $A$V$_3$Sb$_5$ series and may explain not only the enigmatic double superconducting dome in CsV$_3$Sb$_5$ but also the suppressed $T_c$ in (K,Rb)V$_3$Sb$_5$. Our results expand the current understanding of CO in the $A$V$_3$Sb$_5$ series and its relation to other collective phenomena realized in correlated kagome systems.

We start with a brief description of the overall electronic structure of $A$V$_3$Sb$_5$. As displayed in Fig. 2a, the DFT band calculation for CsV$_3$Sb$_5$ reveals four bands near E$_F$: an electron-pocket at the Brillouin zone center Γ (*G*-band), $d_{xy}/d_{x2-y2}$ orbital kagome band with Dirac point at ≈ –0.27 eV and vHS near E$_F$ (*K1*-band), $d_{xz}/d_{yz}$ orbital kagome band with Dirac point at ≈ –1.3 eV and vHS near E$_F$ (*K2*-band), and additional $d_{xz}/d_{yz}$ orbital kagome band with opposite parity from the *K2*-band (*K2'*-band). All band dispersions have been closely reproduced in previous ARPES studies[19,20,36–38]. Meanwhile, we note that in the experimental geometry used in the present study, only *G*-, *K1*-, and *K2*-bands are visible in the ARPES spectra (see Fig. 2c for example) due to the destructive interference of photoelectrons from the *K2'*-band[39]. Additional characterization of the electronic structure in $A$V$_3$Sb$_5$, including the Fermi surfaces, wide-range energy-momentum dispersions, and $k_z$-dependent measurements can be found in the Extended Data Fig. 1 and 2.

As summarized in Fig. 2, we observed two distinct electronic reconstructions induced by 3D-CO in CsV$_3$Sb$_5$. First, as shown in Fig. 2b-d, we detected clear shadow bands below $T_{CO}$

(dashed arrows in Fig. 2d), which are the replica of the original bands (solid arrows in Fig. 2c,d) folded along the in-plane momentum direction. This is a direct consequence of the new periodicity arising from the in-plane component of charge order, which folds the pristine Brillouin zone to the smaller 2×2 CO Brillouin zone (see schematics in Fig. 2b). Such shadow bands and in-plane folding effects have been observed in other charge order systems such as transition metal dichalcogenides[40] and rare-earth trichalcogenides[41]. At the same time, as shown in Fig. 2e-k, a detailed inspection below $T_{CO}$ additionally reveals an unusual doubling or splitting of the kagome bands along the energy axis. Such splitting could be visualized only after careful optimization of the spectral quality[42] (see also Fig. 4f-h for corresponding energy and momentum distribution curves). At the simplest level, one can understand the band doubling as a consequence of the out-of-plane component of the 3D-CO, which folds the Brillouin zone along the $k_z$-direction and superimposes the $k_z = \pi \sim \pi/2$ bands onto the $k_z = 0 \sim \pi/2$ bands (see schematics in Fig. 2e). In the case of CsV$_3$Sb$_5$, we find three regions in the band structure where the doubling becomes most prominent: near the vHS of the $K1$-band (Fig. 2g,k), at the lower Dirac band of the $K1$-band (Fig. 2g,i), and at the $K2$-band near E$_F$ (Fig. 2k). We also refer to the Extended Data Figure 3 for the corresponding energy- and momentum-distribution curves highlighting the band splitting at low temperature.

The key idea of this study comes from the realization that the doubled-band dispersion in the 3D-CO state is actually more than the simple superposition of $k_z = 0 \sim \pi/2$ and $k_z = \pi \sim \pi/2$ bands of the pristine structure. In the 3D-CO state, the adjacent kagome layers in $A$V$_3$Sb$_5$ become distinct upon realizing different CO patterns on each layer (Fig. 1j-m). The altered hopping pathways between the two charge-ordered kagome planes further reconstruct the doubled-band dispersion. This mechanism depends on the detailed ionic displacement patterns in adjacent kagome planes, making the band splitting strongly dependent on the 3D-CO structure. This provides a unique and highly constrained approach to resolve the microscopic structure of 3D-CO in $A$V$_3$Sb$_5$ series.

To illustrate this idea further, we simulated the reconstruction of CsV$_3$Sb$_5$ bands in possible 3D-CO structures using DFT. Fig. 3a-f show the electronic structure of CsV$_3$Sb$_5$ at $k_z = 0$ in the inverse *MMM* (a), *MMM* (b), *LLL+MMM* (c), inverse *MLL* (d), and *MLL* (e) phases (see Table S1 for the exact structures used in the calculation). In accordance with the experimental results, the doubling of the $K1$- and $K2$-kagome bands is closely reproduced in the 3D-CO phases (see yellow arrows in Fig. 3c for example). Crucially, the details of band splitting become noticeably different in each 3D-CO phase, directly reflecting the type of in-plane distortion (TrH vs. SoD) and the stacking order along the c axis (see Supplementary Discussion 2 for intuitive way to understand the splitting in each 3D-CO phase). The most noticeable discriminant between different 3D-CO structures is the behavior of the lower *K1*-Dirac band. As highlighted with the green boxes in Fig. 3, the lower *K1*-Dirac band barely splits in the inverse *MMM*, *MMM*, inverse *MLL*, and *MLL* structures (Fig. 3a,b,d,e), while an apparent doubling is observed in the *LLL+MMM* structure (Fig. 3c). The doubling of the lower *K1*-Dirac band in the latter closely reproduces the ARPES spectra in Fig. 2g,i, and is a direct consequence of the coexistence of two charge order gaps arising from the alternating SoD and TrH layers in the 3D-CO structure (see below and Supplementary Discussion 2). We thus conclude that the observed band splitting supports the *LLL+MMM* structure or Alternating SoD and TrH phase (Fig. 1j) as the microscopic 3D-CO structure in $CsV_3Sb_5$.

Intriguingly, the investigation of $KV_3Sb_5$, $RbV_3Sb_5$, and Sn-doped $CsV_3Sb_5$ ($CsV_3Sb_{4.96}Sn_{0.04}$) revealed an electronic reconstruction markedly different from the $CsV_3Sb_5$ case. Figure 4 displays ARPES spectra of $KV_3Sb_5$, $RbV_3Sb_5$, and Sn-doped $CsV_3Sb_5$ measured at 6 K, in the CO state. Similar to the case of $CsV_3Sb_5$, the doubling or splitting of the *K1*-vHS (Fig. 4a,b,c) and *K2*-bands (Fig. 4d,e) is clearly observed across the whole family. The corresponding energy distribution curves of the *K1*-vHS (Fig. 4f) and momentum distribution curves of the *K2*-band (Fig. 4g) also unambiguously demonstrate the presence of band doubling. We note that, in Fig. 4f, the clear trend of increasing magnitude of splitting with decreasing size of the alkali metal is apparent, and it may reflect the enhanced degree of $k_z$ dispersion from Cs to K.[1,27] We thus conclude that the 3D-CO is a universal phenomenon in the $AV_3Sb_5$ series. However, we observe that the behavior of the lower *K1*-Dirac band in $KV_3Sb_5$, $RbV_3Sb_5$, and Sn-doped $CsV_3Sb_5$ is very different from that of the $CsV_3Sb_5$ case. As highlighted with yellow arrows in Fig. 4a-e, the lower Dirac dispersion of the *K1* band does not undergo a splitting in the 3D-CO state, in contrast to the observations in $CsV_3Sb_5$ (Fig. 2g,i). The absence of the splitting for the lower *K1* Dirac band can be further confirmed from the momentum distribution curves in Fig. 4h. Compared to the calculations in Fig. 3, this behavior rules out the *LLL+MMM* structures (Fig. 3c,f) but is consistent with the inverse *MMM*, *MMM*, inverse *MLL*, and *MLL* structures (Fig. 3a,b,d,e). In case of the inverse *MMM* and *MMM* structures however, the charge distortions are identical in neighboring kagome planes (2×2×1 structure), therefore, they cannot induce the out-of-plane band doubling observed on the $K1$-vHS and $K2$-band. This leaves the inverse $MLL$ or $MLL$ phases, intrinsically breaking $C_6$-rotational symmetry, as possible ionic displacement motifs in KV$_3$Sb$_5$, RbV$_3$Sb$_5$, and Sn-doped CsV$_3$Sb$_5$.

One can further distinguish the inverse $MLL$ (Staggered SoD) and $MLL$ (Staggered TrH) phases by recognizing the difference in the charge order gaps induced by the in-plane SoD and TrH distortions. As shown in the DFT band structures of Fig 4i,j, the TrH distortion of the $MLL$ phase produces a much larger gap at the M point, $\Delta_{TrH} = 0.22$ eV, compared to the SoD distortion, $\Delta_{SoD} = 0.13$ eV. In the case of the $LLL+MMM$ structure with alternating SoD and TrH kagome planes, the SoD and TrH gaps simultaneously manifest in the spectrum (Fig. 4k), resulting in the splitting of the lower $K1$-Dirac band observed in CsV$_3$Sb$_5$. For the detailed inspection of the CO gap near the M point, we acquired additional ARPES maps with vertical light polarization (Fig. 4l-o). The close comparison of the ARPES spectra of KV$_3$Sb$_5$, RbV$_3$Sb$_5$, and Sn-doped CsV$_3$Sb$_5$ (Fig. 4l-n) with that of CsV$_3$Sb$_5$ (Fig. 4o) clearly suggests that the CO gap of the former compounds corresponds to the larger gap, i.e., to a TrH distortion. We thus conclude that the 3D-CO in KV$_3$Sb$_5$, RbV$_3$Sb$_5$, and Sn-doped CsV$_3$Sb$_5$ manifests in the $MLL$ phase at variance with the $LLL+MMM$ structure in pristine CsV$_3$Sb$_5$. This implies that despite the charge ordering tendency of the kagome lattice being universal in the $A$V$_3$Sb$_5$ family, the microscopic details of 3D-CO are strongly dependent on the chemical composition and doping, adding to the rich physics that can be realized in the $A$V$_3$Sb$_5$ series.

The observation of the evolution and tunability of the 3D-CO structure within the $A$V$_3$Sb$_5$ family offers important insights into the exotic electronic phenomena realized in $A$V$_3$Sb$_5$ and their dependence on chemical composition (see Supplementary Discussion 3). In the following, we discuss the empirical correlation between the 3D-CO structure and the superconducting phenomenology of the $A$V$_3$Sb$_5$ family, providing a new framework to understand the double-dome superconductivity in CsV$_3$Sb$_5$ and the suppression of $T_c$ in KV$_3$Sb$_5$ and RbV$_3$Sb$_5$.

One of the most intriguing aspects of the CsV$_3$Sb$_5$ phase diagram is the emergence of a double superconducting dome, which signals a nontrivial interplay between charge order and superconductivity. The double superconducting dome was first discovered in the pressure-temperature phase diagram of CsV$_3$Sb$_5$ (Ref.[43,44]) and subsequently observed as a function of Sn substitution in CsV$_3$Sb$_{5-x}$Sn$_x$ as reproduced in Fig. 5e (Ref.[45]). However, the origin of this behavior has been unclear so far. To construct the corresponding phase diagram of 3D-CO, we investigate the band structure reconstruction as a function of Sn-doping, $x$ = 0, 0.02, 0.03, and 0.04. As summarized in Fig. 5a-d, the *LLL+MMM* phase of the pristine $CsV_3Sb_5$ persists to the $x$ = 0.02 and 0.03 samples as evidenced by the splitting of the lower *K1* Dirac band. The splitting disappears at $x$ = 0.04, signaling the transition of 3D-CO structure to the *MLL* phase as discussed in Fig. 4. Notably, this transition from the *LLL+MMM* to *MLL* phase takes place near the maximum of the first superconducting dome as illustrated in Fig. 5e. This suggests that the transition to the *MLL* phase and its stronger competition with superconductivity might be responsible for the suppression of $T_c$ at intermediate Sn concentrations and provide a basis to understand the emergence of a double superconducting dome. Intriguingly, applying the same argument to $KV_3Sb_5$ and $RbV_3Sb_5$ – which are already in the *MLL* phase in the pristine state (Fig. 5f) – might explain their markedly lower superconducting transition temperature ($T_c$ = 0.92 K and 0.93 K respectively) compared to $CsV_3Sb_5$ ($T_c$ = 2.5 K).[1] We also note that the superconductivity in $(K,Rb)V_3Sb_5$ exhibits only a single dome as a function of both pressure and Sn substitution (Fig. 5f).[46–48] This is again consistent with the absence of the *LLL+MMM* to *MLL* transition in contrast to the case of $CsV_3Sb_5$.

    In conclusion, we establish the microscopic structure of 3D-CO, its tunability, and its potential implications for the superconducting state in the family of topological kagome metals $A$V$_3$(Sb,Sn)$_5$. Determining the exact structure of 3D-CO is a topic of great relevance, as it sets the background symmetry of the system under which other many-body effects emerge. Combining high-resolution ARPES and DFT supercell calculations, we resolve the detailed electronic reconstruction of the kagome bands in the charge order state from which we constrain the microscopic 3D-CO pattern. We reveal that the 3D-CO in $CsV_3Sb_5$ consists of a stacking of kagome layers with alternating SoD and TrH distortions (*LLL+MMM*), while $KV_3Sb_5$, $RbV_3Sb_5$, and Sn-doped $CsV_3Sb_5$ realize the staggered TrH distortions breaking the $C_6$ rotational symmetry (*MLL*). The remarkable tunability of the 3D-CO state across otherwise similar compounds suggests that the $A$V$_3$Sb$_5$ series is a candidate host for an extremely rich phase diagram of emergent electronic phases, enabling new opportunities for fundamental studies at the nexus of strong correlation phenomena and topology.

**Acknowledgements**

This work was supported by the Air Force Office of Scientific Research Young Investigator Program under grant FA9550-19-1-0063, and by the STC Center for Integrated Quantum Materials (NSF grant no. DMR-1231319). The work is funded in part by the Gordon and Betty Moore Foundation's EPiQS Initiative, Grant GBMF9070 to JGC. The works at Max Planck POSTECH/Korea Research Initiative were supported by the National Research Foundation of Korea funded by the Ministry of Science and ICT, Grant No. 2022M3H4A1A04074153 and 2020M3H4A2084417. B.R.O. and S.D.W. were supported by the National Science Foundation (NSF) through Enabling Quantum Leap: Convergent Accelerated Discovery Foundries for Quantum Materials Science, Engineering and Information (Q-AMASE-i): Quantum Foundry at UC Santa Barbara (DMR-1906325). This research used resources of the Advanced Light Source, a U.S. DOE Office of Science User Facility under contract no. DE-AC02-05CH11231. M.K. acknowledges a Samsung Scholarship from the Samsung Foundation of Culture. B.R.O. acknowledges support from the California NanoSystems Institute through the Elings Fellowship program.

**Author contributions**

M.K., J.-H.P, and R.C. conceived the project; M.K. and J.Y. performed the ARPES experiments and analyzed the resulting data with help from S.H.R., J.K., C.J., A.B., and E.L.; S.F. performed the theoretical calculations with help from E.K., J.C.; B.R.O., Y.M.O., and S.D.W. synthesized and characterized the crystals. M.K. and R.C. wrote the manuscript with input from all coauthors.

**Competing interests**

The authors declare no competing interests

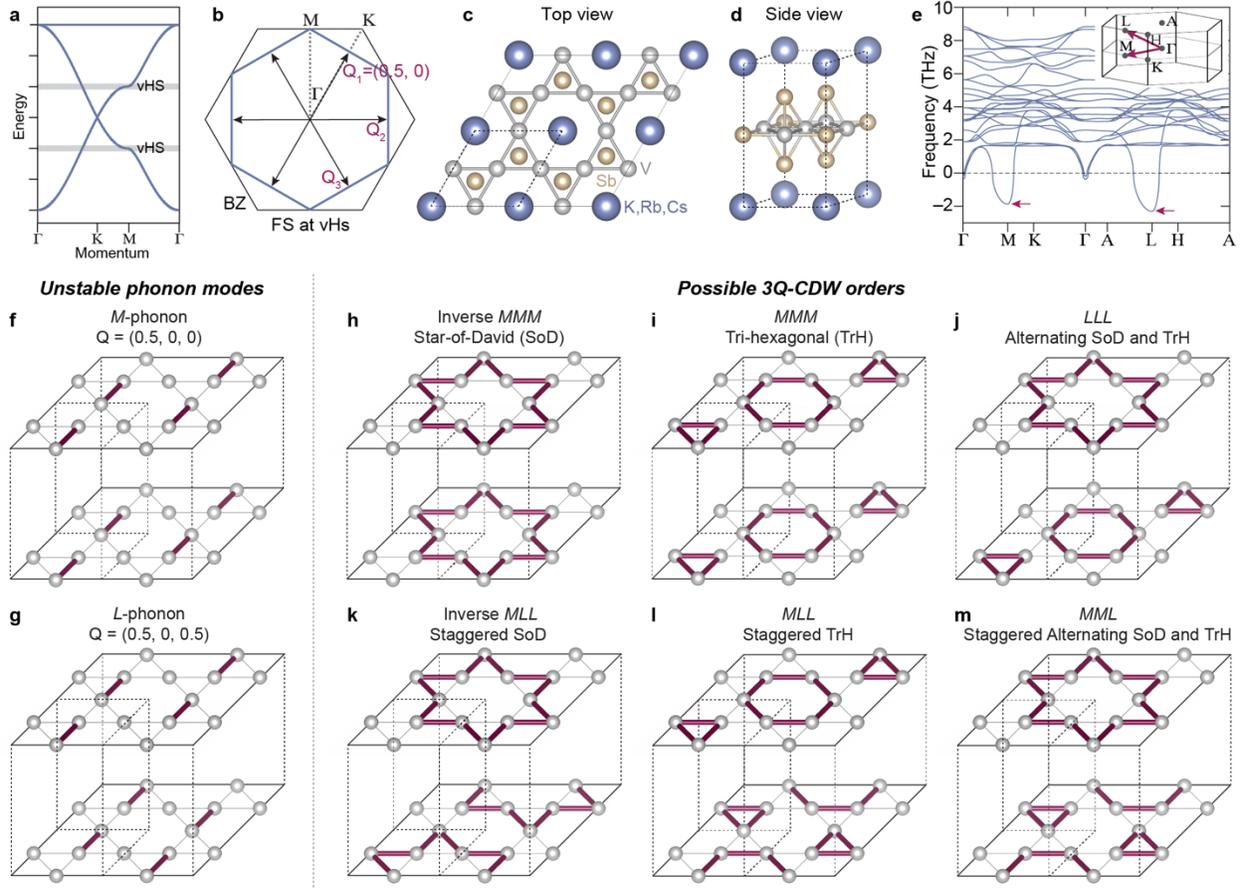

**Figure 1 | Possible microscopic structures of the three-dimensional charge order (3D-CO) in kagome metal $A$V$_3$Sb$_5$. a,** Tight-binding electronic structure of an ideal kagome lattice. Grey-shaded lines mark the vHS filling fractions at 3/12 and 5/12. **b,** Perfectly nested hexagonal Fermi surface of the kagome lattice at the vHS filling fractions of **a**. Double-headed arrows indicate three symmetry-equivalent nesting vectors $Q_1$, $Q_2$, and $Q_3$. **c,d,** Crystal structure of the $A$V$_3$Sb$_5$ with a V-kagome net. Dashed lines mark the unit cell in the undistorted phase. **e,** Calculated phonon dispersion of CsV$_3$Sb$_5$ showing instabilities of the pristine structure at $M$ and $L$. Inset shows the Brillouin zone. **f,g,** Lattice distortions corresponding to the instabilities at $M$ and $L$ phonons, respectively. **h-m,** Possible structures of the 3D-CO in $A$V$_3$Sb$_5$ based on 3$Q$-combinations of the $M$ and $L$ phonons.

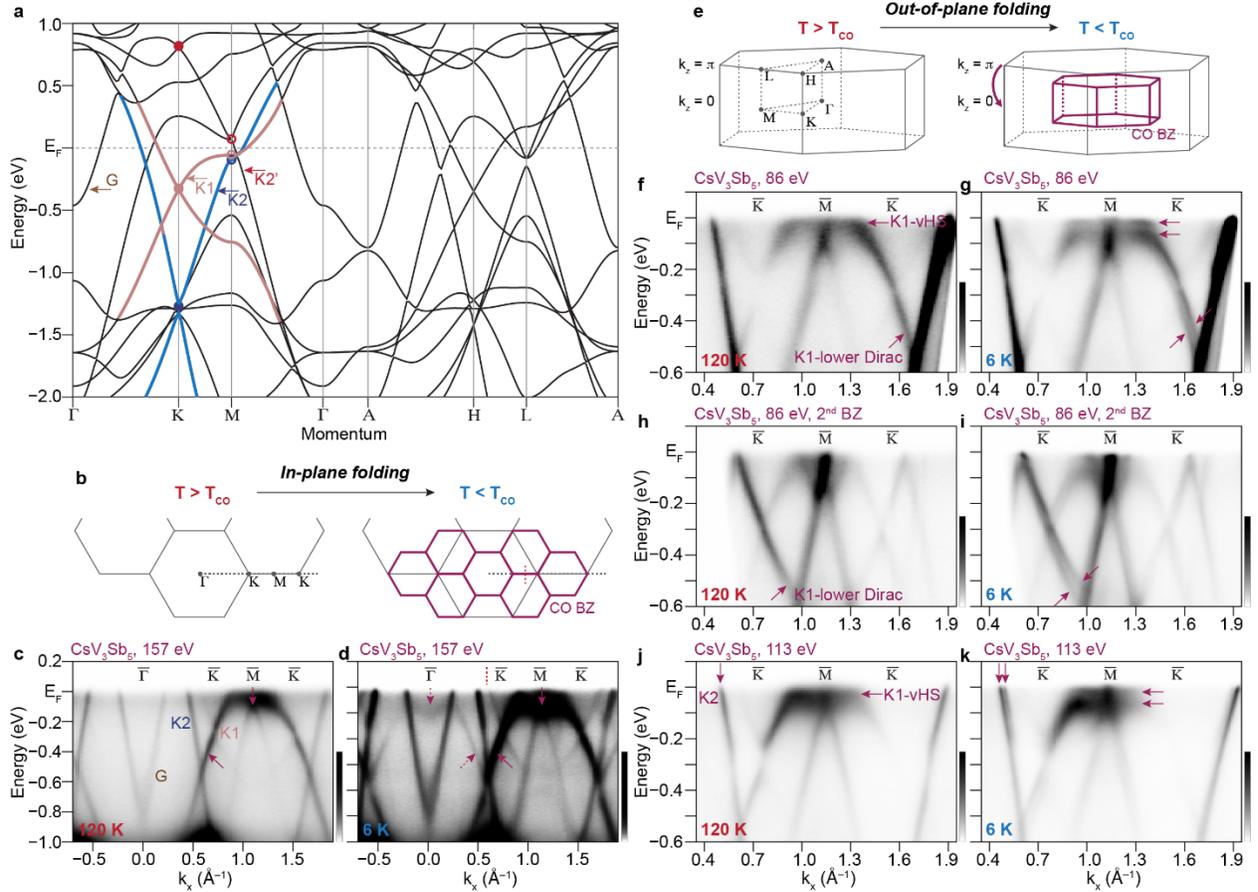

**Figure 2 | Two distinct types of electronic reconstructions in CsV$_3$Sb$_5$ induced by 3D-CO. a,** DFT band structure of CsV$_3$Sb$_5$ showing four bands crossing the Fermi level: *G*, *K1*, *K2*, and *K2'*-bands. The Dirac points at *K* and vHS at *M* emerging from the *K1*, *K2*, and *K2'* kagome bands are marked with filled and open circles, respectively. Thick coral and blue lines highlight the *K1*- and *K2*-bands actively discussed in the ARPES data. **b-d,** Electronic reconstruction from the in-plane component of charge order. **b,** Schematics of the in-plane folding of the Brillouin zone. **c,d,** Experimental band dispersion of CsV$_3$Sb$_5$ measured at 120 K and 6 K (above and below T$_{CO}$), respectively. Solid arrows in c,d mark the original bands, while the dashed arrows in d indicate the replica bands. **e-k,** Electronic reconstruction from the out-of-plane component of 3D-CO. **e,** Schematics of the out-of-plane folding of the Brillouin zone. Panels f,h,j (g,i,k) represent the dispersions measured above (below) T$_{CO}$, at the first Brillouin zone with photon energy 86 eV, at the second Brillouin zone with photon energy 86 eV, and at the first Brillouin zone with photon energy 113 eV, respectively. Solid arrows in g,i,k indicate the doubling or splitting of the kagome bands in the 3D-CO state. See also Supplementary Figure S2 for the visualization of splitting at other photon energies.

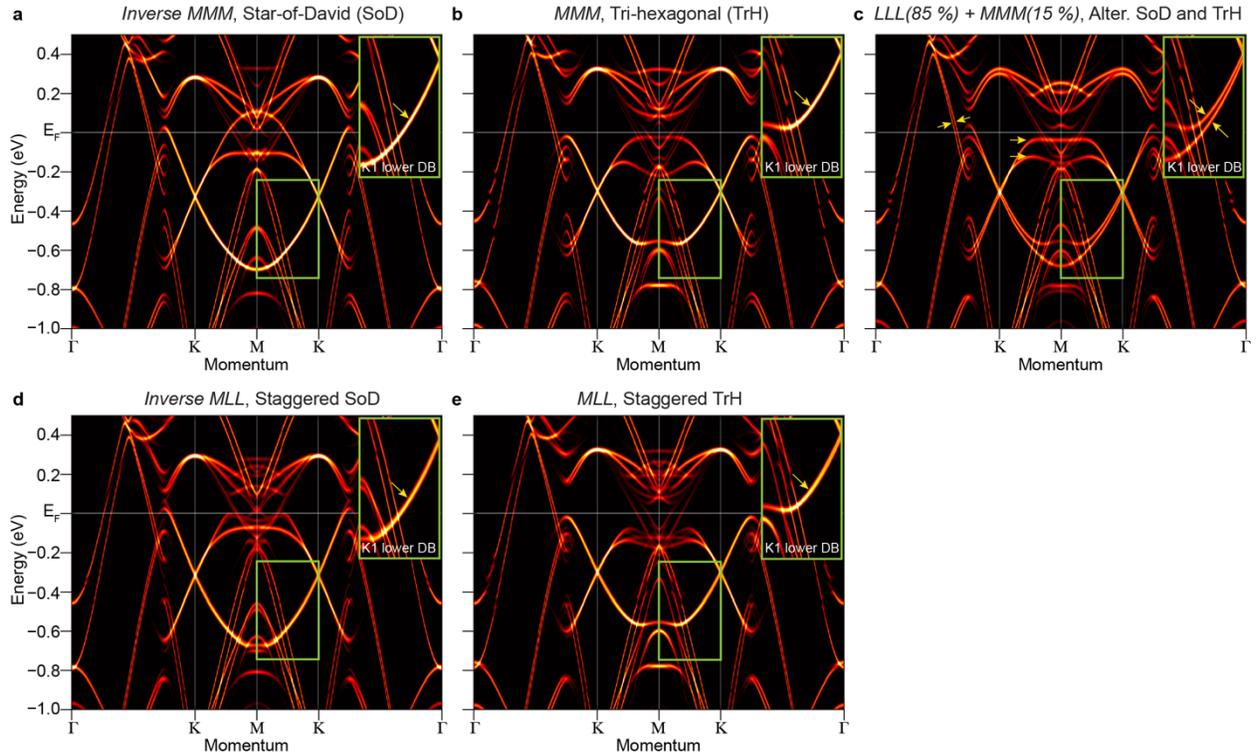

**Figure 3 | Theoretical calculation of electronic reconstruction in $A$V$_3$Sb$_5$ and its dependence on the microscopic structure of 3D-CO. a-f,** Calculated dispersion of CsV$_3$Sb$_5$ at $k_z = 0$ in the inverse *MMM*, *MMM*, *LLL+MMM*, inverse *MLL*, and *MLL* phases, respectively. For simplicity, the dispersion is unfolded along the in-plane momentum direction. Yellow arrows in c highlight the splitting of *K1* and *K2* bands in the 3D-CO state. For the inverse *MLL* and *MLL* (d,e) the calculation is averaged over three $C_2$ symmetric charge order domains to account for the macroscopic beam spot size. Corresponding domain-resolved dispersions are presented in the Extended Data Fig. 4. For the inverse *MMM* (a) and *MMM* (b) structures with the 2×2×1 periodicity, we artificially folded the Brillouin zone along $k_z$ for the proper comparison with other structures. Insets: zoomed-in view of the lower *K1* Dirac band, whose splitting sensitively depends on the microscopic structure of 3D-CO (see yellow arrows).

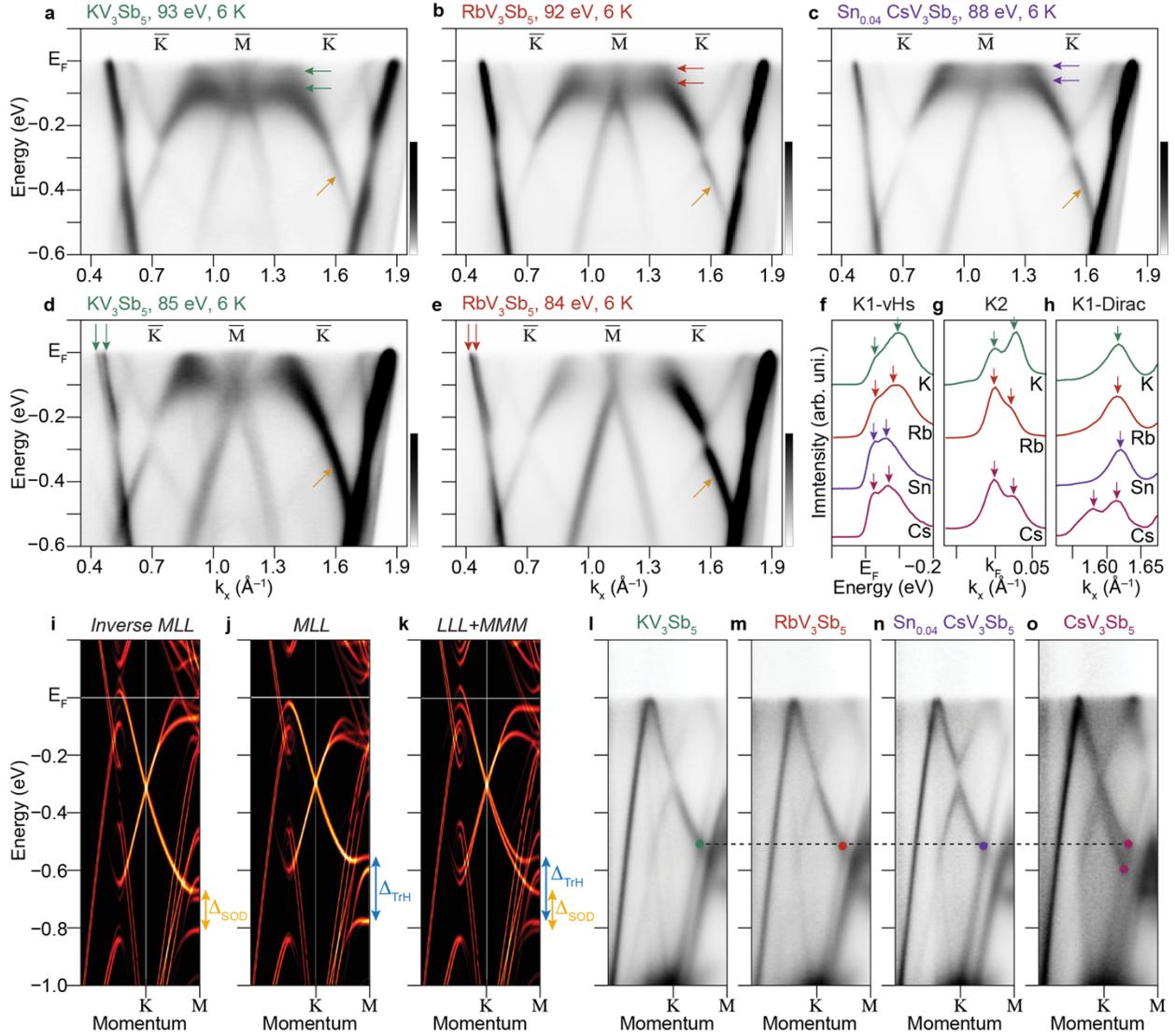

**Figure 4 | Electronic reconstructions in $KV_3Sb_5$, $RbV_3Sb_5$, and Sn-doped $CsV_3Sb_5$ in 3D-CO state. a-e,** ARPES spectra of $KV_3Sb_5$ (a,b) $RbV_3Sb_5$ (c,d), and Sn-doped $CsV_3Sb_5$ (e) measured at 6 K. The spectra in a-e are collected with 93 eV, 85 eV, 92 eV, 84 eV, and 88 eV photons, respectively. Green, red, and purple arrows indicate the splitting of *K1*-vHS and *K2*-band. Yellow arrows highlight the absence of splitting on the lower Dirac band of *K1*. **f,** EDCs measured near the Fermi momentum ($k_F$) of the *K1*-vHs. **g,** MDCs of the *K2*-band measured at the Fermi energy ($E_F$). **h,** MDCs of the lower *K1*-Dirac band measured at –0.4 eV. The arrows in f-h highlight the presence or absence of the 3D-CO induced splitting. **i-k,** Comparison between the DFT band structures of inverse *MLL*, *MLL*, and *LLL+MMM* phases. Double-headed arrows indicate the charge order gaps at the M point induced by the SoD and TrH distortions. **l-o,** Comparison between the ARPES spectra of $KV_3Sb_5$, $RbV_3Sb_5$, Sn-doped $CsV_3Sb_5$, and $CsV_3Sb_5$ taken with vertically polarized light. The dashed line and filled circles are guides to eye emphasizing that the charge order gaps in $KV_3Sb_5$, $RbV_3Sb_5$, Sn-doped $CsV_3Sb_5$ correspond to the TrH distortion of the *MLL* phase.

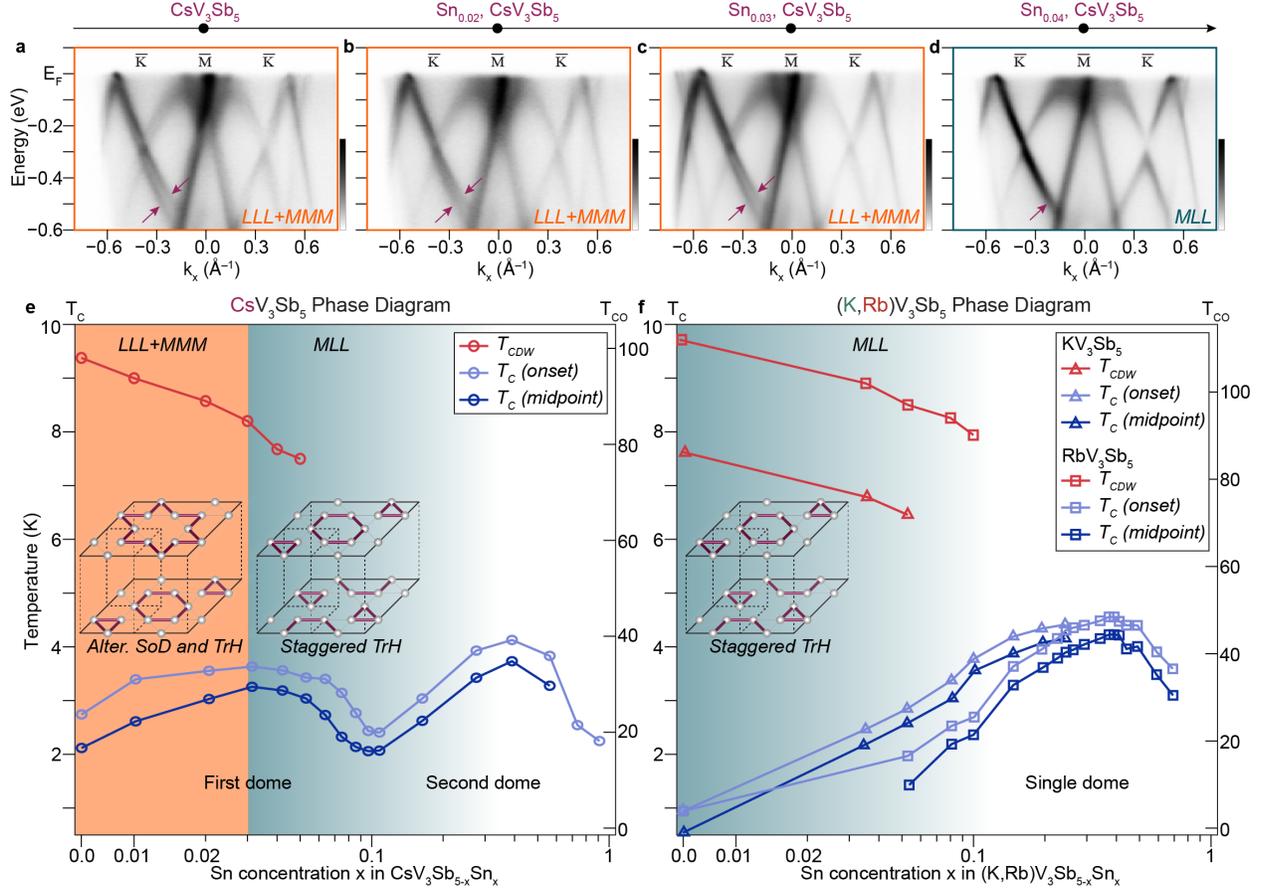

**Figure 5 | Phase diagram of the 3D-CO in CsV$_3$Sb$_{5-x}$Sn$_x$ and its correlation with double-dome superconductivity. a-d,** Evolution of the electronic reconstruction in CsV$_3$Sb$_{5-x}$Sn$_x$ as a function of Sn-doping x = 0, 0.02, 0.03, and 0.04, respectively. The purple arrows highlight the presence (a-c) or absence (d) of the splitting on the lower K1 Dirac band, marking the transition of the 3D-CO structure from *LLL+MMM* to the *MLL* phase. **e,** Phase diagram of CsV$_3$Sb$_{5-x}$Sn$_x$. The open blue (light-blue) circles represent the superconducting transition temperature estimated from the midpoint (onset) of the transition. The red circles represent the charge order transition temperature $T_{CO}$. Both $T_c$ and $T_{CO}$ are reproduced from Ref.[45] The orange and teal background represents the region with *LLL+MMM* and *MLL* phases, respectively, as also shown in the insets. **f,** Corresponding phase diagram of (K,Rb)V$_3$Sb$_5$ with only *MLL* 3D-CO phase and single superconducting dome. Both $T_c$ and $T_{CO}$ are adopted from Ref.[47]

**References**

1. Ortiz, B. R. *et al.* New kagome prototype materials: Discovery of KV3Sb5,RbV3Sb5, and CsV3Sb5. *Phys. Rev. Mater.* **3**, 94407 (2019).
2. Ortiz, B. R. *et al.* CsV3Sb5: a Z2 topological kagome metal with a superconducting ground state. *Phys. Rev. Lett.* **125**, 247002 (2020).
3. Keimer, B., Kivelson, S. A., Norman, M. R., Uchida, S. & Zaanen, J. From quantum matter to high-temperature superconductivity in copper oxides. *Nature* **518**, 179–186 (2015).
4. Si, Q., Yu, R. & Abrahams, E. High-temperature superconductivity in iron pnictides and chalcogenides. *Nat. Rev. Mater.* **1**, 1 (2016).
5. Cao, Y. *et al.* Unconventional superconductivity in magic-angle graphene superlattices. *Nature* **556**, 43–50 (2018).
6. Jiang, Y.-X. *et al.* Unconventional chiral charge order in kagome superconductor KV3Sb5. *Nat. Mater.* **20**, 1353–1357 (2021).
7. Zhao, H. *et al.* Cascade of correlated electron states in a kagome superconductor CsV3Sb5. *Nature* **599**, 216–221 (2021).
8. Shumiya, N. *et al.* Intrinsic nature of chiral charge order in the kagome superconductor Rb V3Sb5. *Phys. Rev. B* **104**, 035131 (2021).
9. Li, H. *et al.* Rotation symmetry breaking in the normal state of a kagome superconductor KV3Sb5. *Nat. Phys.* (2022) doi:10.1038/s41567-021-01479-7.
10. Xiang, Y. *et al.* Twofold symmetry of c-axis resistivity in topological kagome superconductor CsV3Sb5 with in-plane rotating magnetic field. *Nat. Commun.* **12**, 6727 (2021).
11. Yu, L. *et al.* Evidence of a hidden flux phase in the topological kagome metal CsV$_3$Sb$_5$. *arXiv* 2107.10714 (2021).
12. Ortiz, B. R. *et al.* Superconductivity in the Z2 kagome metal KV3Sb5. *Phys. Rev. Mat.* **5**, 034801 (2021).
13. Yin, Q. *et al.* Superconductivity and normal-state properties of kagome metal RbV3Sb5 single crystals. *Chinese Phys. Lett.* **38**, 037403 (2021).
14. Wu, X. *et al.* Nature of Unconventional Pairing in the Kagome Superconductors A V 3 Sb 5 ( A = K , Rb , Cs ). *Phys. Rev. Lett.* **127**, 177001 (2021).
15. Tazai, R., Yamakawa, Y., Onari, S. & Kontani, H. Mechanism of exotic density-wave and beyond-Migdal unconventional superconductivity in kagome metal AV 3 Sb 5 (A=K, Rb, Cs). *Arxiv* 2107.05372 (2021).
16. Zhou, X. *et al.* Origin of charge density wave in the kagome metal CsV3 Sb5 as revealed by optical spectroscopy. *Phys. Rev. B* **104**, L041101 (2021).
17. Li, H. *et al.* Observation of Unconventional Charge Density Wave without Acoustic Phonon Anomaly in Kagome Superconductors A V3Sb5 (A=Rb, Cs). *Phys. Rev. X* **11**, 031050 (2021).
18. Nakayama, K. *et al.* Multiple energy scales and anisotropic energy gap in the charge-density-wave phase of the kagome superconductor CsV 3 Sb 5. *Phys. Rev. B* **104**, L161112 (2021).
19. Liu, Z. *et al.* Charge-Density-Wave-Induced Bands Renormalization and Energy Gaps in a Kagome Superconductor RbV 3 Sb 5 . *Phys. Rev. X* **11**, 41010 (2021).
20. Kang, M. *et al.* Twofold van Hove singularity and origin of charge order in topological kagome superconductor CsV3Sb5. *Nat. Phys.* (2022) doi:10.1038/s41567-021-01451-5.

21. Guo, H. M. & Franz, M. Topological insulator on the kagome lattice. *Phys. Rev. B* **80**, 113102 (2009).
22. Xu, G., Lian, B. & Zhang, S.-C. Intrinsic Quantum Anomalous Hall Effect in the Kagome Lattice Cs2LiMn3F12. *Phys. Rev. Lett.* **115**, 186802 (2015).
23. Tang, E., Mei, J.-W. & Wen, X.-G. High-Temperature Fractional Quantum Hall States. *Phys. Rev. Lett.* **106**, 236802 (2011).
24. Yu, S. L. & Li, J. X. Chiral superconducting phase and chiral spin-density-wave phase in a Hubbard model on the kagome lattice. *Phys. Rev. B* **85**, 4–7 (2012).
25. Kiesel, M. L., Platt, C. & Thomale, R. Unconventional fermi surface instabilities in the kagome hubbard model. *Phys. Rev. Lett.* **110**, 126405 (2013).
26. Wang, W., Li, Z., Xiang, Y. & Wang, Q. Competing electronic orders on kagome lattices at van Hove filling. *Phys. Rev. B* **87**, 115135 (2013).
27. Tan, H., Liu, Y., Wang, Z. & Yan, B. Charge density waves and electronic properties of superconducting kagome metals. *Phys. Rev. Lett.* **127**, 046401 (2021).
28. Christensen, M. H., Birol, T., Andersen, B. M. & Fernandes, R. M. Theory of the charge density wave in AV3Sb5 kagome metals. *Phys. Rev. B* **104**, 214513 (2021).
29. Subedi, A. Hexagonal-to-base-centered-orthorhombic 4Q charge density wave order in kagome metals KV3Sb5, RbV3Sb5, and CsV3Sb5. *Phys. Rev. Mater.* **6**, 015001 (2022).
30. Liang, Z. *et al.* Three-Dimensional Charge Density Wave and Surface-Dependent Vortex-Core States in a Kagome Superconductor CsV3Sb5. *Phys. Rev. X* **11**, 31026 (2021).
31. Ortiz, B. R. *et al.* Fermi Surface Mapping and the Nature of Charge-Density-Wave Order in the Kagome Superconductor CsV3Sb5. *Phys. Rev. X* **11**, 41030 (2021).
32. Li, H. *et al.* Spatial symmetry constraint of charge-ordered kagome superconductor.
33. Ratcliff, N., Hallett, L., Ortiz, B. R., Wilson, S. D. & Harter, J. W. Coherent phonon spectroscopy and interlayer modulation of charge density wave order in the kagome metal CsV3Sb5. *Phys. Rev. Mater.* **5**, L111801 (2021).
34. Luo, J. *et al.* Star-of-David pattern charge density wave with additional modulation in the kagome superconductor CsV$_3$Sb$_5$ revealed by $^{51}$V-NMR and $^{121/123}$Sb-NQR. *arXiv* 2108.10263 (2021).
35. Ye, Z., Luo, A., Yin, J.-X., Hasan, M. Z. & Xu, G. Structural instability and charge modulations in the Kagome superconductor $A$V$_3$Sb$_5$. *arXiv* 2111.07314 (2021).
36. Hu, Y. *et al.* charge order assisted topological surface states and flat bands in the kagome superconductor CsV3Sb5. *Sci. Bull.* (2022).
37. Luo, H. *et al.* Electronic Nature of Charge Density Wave and Electron-Phonon Coupling in Kagome Superconductor KV$_3$Sb$_5$. *Nat. Commun.* **13**, 273 (2022).
38. Cho, S. *et al.* Emergence of new van Hove singularities in the charge density wave state of a topological kagome metal RbV3Sb5. *Phys. Rev. Lett.* **127**, 236401 (2021).
39. Hu, Y. Rich Nature of Van Hove Singularities in Kagome Superconductor CsV3Sb5 Yong. *arXiv* 2106.05922 (2021).
40. Rossnagel, K. On the origin of charge-density waves in select layered transition-metal dichalcogenides. *J. Phys. Condens. Matter* **23**, (2011).
41. Brouet, V. *et al.* Angle-resolved photoemission study of the evolution of band structure and charge density wave properties in RTe3. *Phys. Rev. B* **77**, 235104 (2008).
42. Hu, Y. *et al.* Coexistence of Tri-Hexagonal and Star-of-David Pattern in the Charge Density Wave of the Kagome Superconductor AV3Sb5. *arXiv* 2201.06477 (2022).
43. Chen, K. Y. *et al.* Double superconducting dome and triple enhancement of T c in the

kagome superconductor CsV$_3$Sb$_5$ under high pressure. *Phys. Rev. Lett.* **126**, 247001 (2021).
44. Yu, F. H. *et al.* Unusual competition of superconductivity and charge-density-wave state in a compressed topological kagome metal. *Nat. Commun.* **12**, 10–15 (2021).
45. Oey, Y. M. *et al.* Fermi level tuning and double-dome superconductivity in the kagome metals CsV$_3$Sb$_{5-x}$Sn$_x$. *Phys. Rev. Mat.* **6**, 041801 (2022).
46. Du, F. *et al.* Pressure-induced double superconducting domes and charge instability in the kagome metal KV3Sb5. *Phys. Rev. B* **103**, L220504 (2021).
47. Zhu, C. C. *et al.* Double-dome superconductivity under pressure in the V-based Kagome metals AV3Sb5 (A = Rb and K). **094507**, 5–9 (2021).
48. Oey, Y. M., Kaboudvand, F., Ortiz, B. R., Seshadri, R. & Wilson, S. D. Tuning charge-density wave order and superconductivity in the kagome metals KV$_3$Sb$_{5-x}$Sn$_x$ and RbV$_3$Sb$_{5-x}$Sn$_x$. *arXiv* 2205.06317 (2022).

# Methods

## Sample synthesis and angle-resolved photoemission spectroscopy

High-quality single crystals of pristine and Sn-doped $A$V$_3$Sb$_5$ were synthesized via flux method as described in Ref.[1,2,45]. ARPES experiments were performed at Beamline 7.0.2 (MAESTRO) of the Advanced Light Source, equipped with R4000 hemispherical electron analyzer (Scienta Omicron). The samples were cleaved inside an ultrahigh vacuum chamber with a base pressure better than $\approx 4 \times 10^{-11}$ torr. We keep the following experimental geometry throughout the measurement except otherwise specified: horizontal analyzer slit, linear horizontal light polarization, and Γ-K-M direction of the sample aligned to the scattering plane. For each sample, photon energy was scanned from 60 eV to 200 eV, covering more than three complete three-dimensional Brillouin zone. The $k_z$ plane is determined based on the nearly-free-electron final state approximation. For the high-resolution data in Fig. 2,4, we selected the photon energies for each sample that best visualize the band splitting around $k_z \approx 0$. The energy and momentum resolutions were better than 20 meV and 0.01 Å$^{-1}$.

## Density functional theory calculations

DFT calculations were performed using the Vienna Ab initio Simulation Package[49,50], with GGA-PBE exchange-correlation functional[51] and the pseudo potential formalism based on the Projector Augmented Wave method[52]. The phonon frequency spectrum was derived from the Hessian matrix (which encodes the second derivatives of the atomic position coordinate) computed with the density-functional-perturbation theory (DFPT) method, using a 4×4×2 supercell unit[53]. We have further simulated various 2×2×2 CO states from combinations of the unstable $M$ and $L$ phonon modes. The CO structures were relaxed with a 350 eV energy cutoff for the plane-wave basis and a 4×4×2 grid sampling in the momentum space Brillouin zone. See Supplementary Table S1 for the detailed characterization of the 3D-CO structures after relaxation. To elucidate the electronic properties of these COs, we performed electronic band structure unfolding and projections based on the Wannier models derived from DFT ground states, using Wannier 90 code.

**Data availability**

Data associated with this paper is available on the Harvard Dataverse at doi: https://doi.org/10.7910/DVN/KJRGXU

**Methods only references**

49. Kresse, G. & Furthmüller, J. Efficiency of ab-initio total energy calculations for metals and semiconductors using a plane-wave basis set. *Comput. Mater. Sci.* **6**, 15–50 (1996).
50. G. Kresse & Furthmu¨ller, J. Efficient iterative schemes for ab initio total-energy calculations using a plane-wave basis set. *Phys. Rev. B* **54**, 11169–11186 (1996).
51. Perdew, J. P., Burke, K. & Ernzerhof, M. Generalized gradient approximation made simple. *Phys. Rev. Lett.* **77**, 3865–3868 (1996).
52. Blochl, P. E. Projector augmented-wave method. *Phys. Rev. B* **50**, 17953–17979 (1994).
53. Togo, A. & Tanaka, I. First principles phonon calculations in materials science. *Scr. Mater.* **108**, 1–5 (2015).